\documentclass[12pt]{article}
\usepackage{amsfonts}
\usepackage{amssymb}
\usepackage{amsbsy}
\usepackage{mathrsfs}
\usepackage{amsmath}
\usepackage{epsfig}      
\usepackage{color}       

\newlength{\TZ}
\newcommand{\BEQ}{\begin{equation}}     
\newcommand{\BEA}{\begin{eqnarray}}
\newcommand{\BD}{\begin{displaymath}}
\newcommand{\EEQ}{\end{equation}}       
\newcommand{\EEA}{\end{eqnarray}}
\newcommand{\ED}{\end{displaymath}}
\newcommand{\bb}{\begin{eqnarray}}
\newcommand{\ee}{\end{eqnarray}}

\newcommand{\pr}{{\rm Pr}}

\newcommand{\D}{{\rm d}}                
\newcommand{\II}{{\rm i}}               
\newcommand{\erfc}{{\rm erfc\,}}        
\newcommand{\erf}{{\rm erf\,}}          

\newcommand{\demi}{\frac{1}{2}}         

\newcommand{\wit}[1]{\widetilde{#1}}    

\renewcommand{\vec}[1]{\boldsymbol{#1}} 

\newcommand{\appsection}[2]{\setcounter{equation}{0}\setcounter{subsection}{0}
\section*{Appendix #1. #2}
\renewcommand{\theequation}{#1.\arabic{equation}}
              \renewcommand{\thesection}{#1}
              \renewcommand{\thefigure}{#1\arabic{figure}}\setcounter{figure}{0} }

\catcode`\@=11
\def\numberbysection{\@addtoreset{equation}{section}
        \def\theequation{\thesection.\arabic{equation}}}
\numberbysection

\definecolor{gruen}{rgb}{0,0.625,0}       
\definecolor{rot}{rgb}{0.75,0,0}          
\definecolor{blau}{rgb}{0,0,0.75}         
\definecolor{casta}{rgb}{0.45,0.20,0}     
\definecolor{gelb}{rgb}{0.825,0.725,0.0}  

\newcommand{\BLAU}[1]{\textcolor{black}{{\rm #1}}}	

\begin{document}

\begin{titlepage}

\vskip 1.5 cm
\begin{center}
{\Large \bf Exact solution of the Glauber-Ising model on the finite-length semi-open chain} 
\end{center}

\vskip 2.0 cm
\centerline{ {\bf Malte Henkel}$^{a,b}$
}
\vskip 0.5 cm
\begin{center}
$^a$Laboratoire de Physique et Chimie Th\'eoriques (CNRS UMR 7019),\\  Universit\'e de Lorraine Nancy,
B.P. 70239, F -- 54506 Vand{\oe}uvre l\`es Nancy Cedex, France\\~\\
$^b$Centro de F\'{i}sica Te\'{o}rica e Computacional, Universidade de Lisboa, \\Campo Grande, P--1749-016 Lisboa, Portugal\\~\\
\end{center}

\begin{abstract}
The exact time-space correlation function of the $1D$ Glauber-Ising model, quenched to temperature $T=0$ and on a semi-open lattice of finite size $N$, is obtained. 
This also allows to deduce the exact empty-interval probability of the dual $1D$ coagulation-diffusion process on a periodic finite ring and to reproduce the 
long-time decay of the particle concentration. These results are consistent with the generic expectations of dynamical finite-size scaling theory. ~\\~\\
\centerline{\textcolor{gruen}{\large \today}~~~}
\end{abstract}

\vfill

\end{titlepage}

\setcounter{footnote}{0}

\section{Introduction} \label{sec1}

Physical {\em ageing phenomena} \cite{Arce21,Vinc24} may arise in a many-body system after a quench, typically from a disordered initial state, 
either onto a critical point where at least two physical phases become indistinguishable or else
into a phase co-existence region where two macroscopic physical phases coexist. In both cases, the after-quench dynamics is a
slow one, which may come from the effects of the critical-point fluctuations or else from the competition between the relaxation 
towards at least two distinct physical states. Microscopically, the system separates into many (correlated or ordered) clusters 
whose mean size $\ell(t)$ is growing with time. The phenomenology of physical ageing is contained in its three defining properties, namely \cite{Stru78}
\begin{enumerate}
\item slow dynamics (relaxations are slower than might be described by simple exponentials)
\item absence of time-translation-invariance
\item dynamical scaling
\end{enumerate}
and these are reflected in the properties of the (coarse-grained) order-parameter
$\phi=\phi(t,\vec{r})$ which depends on the time $t$ and the space coordinates $\vec{r}$. 
For a disordered initial state, the average order-parameter vanishes, viz.
$\langle \phi(t,\vec{r})\rangle=0$. One of the most frequently studied observables is the 
{\em correlation function}\footnote{It\BLAU{s} Fourier transform, the {\em structure function}, can be measured in scattering experiments.} 
\BEQ \label{1.1}
C(t;\vec{r})   = \left\langle \phi(t,\vec{r})\phi(t,\vec{0})\right\rangle  = t^{-b} F_C\left( \frac{|\vec{r}|}{t^{1/z}}\right)
\EEQ
whose scaling is specified here for an algebraically growing domain size $\ell(t)\sim t^{1/z}$ and which then defines the {\em dynamical exponent} $z$. 
The exponent $b$ is an ageing exponent, but in what follows we shall restrict \BLAU{ourselves} to cases where $b=0$. 
For a non-conserved order-parameter and short-ranged interactions, 
one has $z=2$ \cite{Bray94a,Bray94b,Rute95}.\footnote{For a conserved order-parameter, 
one speaks of {\em phase separation} and $z$ takes different values \cite{Bray94b,Rute95}. 
Long-range interactions lead to further modifications \cite{Bray94b,Corb19a,Corb19b,Chris20,Chris21,Muel24}.} 
Up to metric scale factors, the form of the scaling function $F_C$ is generically expected to be universal, 
hence independent of most of the `details' of the underlying microscopic physics,
see \cite{Bray94a,Cugl03,Puri09,Henk10,Taeu14} for reviews. 
The theoretical task of finding the form of $F_C$ is also of practical importance since {\it a priori} 
knowledge of $F_C$ permits long-time predictions on the basis of short-time data. 

The universality of functions such as $F_C$ permits their study via well-chosen and mathematically extremely simplified models. 
In this context, the celebrate\BLAU{d} $1D$ {\em Ising model} continues to play an important r\^ole. 
It is defined on a chain $\Lambda \subset \mathbb{Z}$ with spin variables $\sigma_n=\pm 1$ attached to each of its sites. 
At equilibrium, it is specified through the Hamiltonian \cite{Lenz20,Ising25}
\BEQ \label{1.2}
{\cal H} = - \sum_{n\in\Lambda} \sigma_{n} \sigma_{n+1}
\EEQ
where the exchange coupling was normalised to unity. \BLAU{Because of its short-ranged interactions, in one dimension there is no equilibrium phase transition at any non-vanishing
temperature $T>0$ \cite{Ising25}.} We shall be interested here in some of its non-equilibrium dynamical properties.\footnote{\BLAU{Slow Glauber-Ising dynamics 
in a ferrimagnetic chain made from Co$^+$ ions and organic radical spins strongly antiferromagnetically coupled was studied experimentally, e.g. \cite{Cane02,Pini07,Cegl21,Folt23}.}}
Such a dynamics, at temperature $T$, may be created in a heat-bath formulation by selecting at each time step
$\Delta t$ randomly a site $n\in\Lambda$, whose spin $\sigma_n$ is updated according to the {\em Glauber rule} \cite{Glau63}, with the rate \cite{Godr00a} 
\BEQ \label{1.3}
\sigma_n(t) \mapsto \pm 1 \mbox{\rm\small ~~~\BLAU{with probability} $\frac{1}{2}\left( 1 \pm \tanh \frac{\sigma_{n-1}(t) + \sigma_{n+1}(t)}{T}\right)$}
\EEQ
\BLAU{Clearly, the temperature $T$ is a property of the heat bath. 
In one spatial dimension, Glauber's rule has the remarquable and attractive feature that local spin-observables such as local magnetisation, local correlators and so on 
satisfy closed equations of motion which can be studied analytically, rather than infinite uncoupled hierarchies of equations of motion \cite{Kreu86} which arise generically. Specifically, on}
a discrete chain, the single-time correlator is $C_n(t) := \left\langle \sigma_{n}(t)\sigma_{0}(t)\right\rangle$
where the average is over the thermal histories defined by eq.~(\ref{1.3}). 
The correlator obeys the (rescaled) equation of motion \cite{Glau63,Godr00a,Lipp00,Maye04,Henk04}, with
the short-hand $0\leq \gamma=\tanh(2/T)\leq 1$
\BEQ \label{1.4}
\partial_t C_n(t) = -2 C_n(t) + \gamma \bigl( C_{n-1}(t) + C_{n+1}(t) \bigr) \mbox{\rm\small ~~~when $n\ne 0$~~} \;\; , \;\; C_0(t) = 1 
\EEQ
\BLAU{which only contains correlators $C_n(t)$ at different sites but does not contain any reference to higher multi-point correlators.}
For an initially disordered system, one has $C_n(0)=\delta_{n,0}$ \cite{Glau63,Godr00a}. 

Various aspects of this model have been thoroughly analysed many times 
\BLAU{\cite{Glau63,Feld71,Schu93,Alca94,Lusc96,Bray97,Prad97,Godr00a,Lipp00,Schu01,Maye03,Maye04,Maye05,Henk04,Alie09,Krap10,Verl11,Godr11,Mont13,Godr22}},
notably for $T=0$ \BLAU{where updates which would lead to an increase in $\cal H$ are forbidden.}
\BLAU{Because of the competition between the two stationary and absorbing states where all spins are either $+1$ or $-1$,} 
the \BLAU{$T=0$}-dynamics becomes slow and obeys dynamical scaling with the exponent 
$z=2$.\footnote{\BLAU{A rigorous new bound on the spectral gap implies for all $T\leq T_c$ the improved bound $z\geq 2$ \cite{Masa25}.}} 
In doing so, a main issue is the appropriate treatment of the constraint $C_0(t)=1$, which precludes the immediate application of Fourier series. 
A recently introduced possibility to treat this uses spatial symmetry properties \cite{Durang10,Henk25b}, which provides a convenient way to treat the model in a finite and periodic lattice. 
Here we shall consider how to extend this idea to a chain which is open on one end (following an idea from \cite{Glau63}), find the exact correlation function in this case
and shall analyse how the finite-size effects will modify the scaling (\ref{1.1}) of the infinite-size system, \BLAU{especially for a case when spatial translation-invariance does not hold}. 
This also will allow to study the influence of the boundary conditions
on the form of the scaling function. One of the aims of this study is to provide an explicitly worked out case to serve as a background for more generic studies. One issue
will be how to insert the results to be obtained in the context of finite-size scaling \cite{Fish71,Barb83}, especially for dynamics \cite{Suzuki77}. 

Another aspect of this problem arises from the link of the Glauber-Ising model with stochastic reaction-diffusion processes \cite{Sigg77}. 
Here we shall focus on the $1D$ coagulation-diffusion process, and mainly studied via the so-called {\em empty-interval method}
e.g. \cite{Tous83,benA90,Doer90,Alca94,Krebs94a,Krebs94b,Simo95,Henk95,Hinr97,benA00,Schu01,Krap10,Durang10,Durang11,Afza11,Fort14,Cram17,Shap18,Turb18,Li25,Zahr26}. 
Each site of the lattice is either empty or occupied by a single particle $A$. 
Particles can randomly hop to a nearest-neighbour site and if that site already happens to be occupied, the
two particles undergo, with probability one, a coagulation reaction $A+A\to A$. 
In one dimension, the particle concentration decays\footnote{For brief overviews on experimental results in $1D$ we refer to \cite{Shap18,Li25} and refs. therein.} 
for long times as $c(t)\sim t^{-1/2}$ \cite{Tous83}, which makes it
(i) a slow process (of which in principle the ageing can be analysed) with anomalous transport and 
(ii) is distinct from mean-field theories which hold for dimensions $d>2$ and would give $c_{\rm MF}(t)\sim t^{-1}$ \cite{Henk09}. 
It is long-established that this process is dual to the Glauber-Ising model at $T=0$ \cite{Sigg77,Sant97}. 
We shall be interested in extending this to finite lattices, with $N$ sites. 
\BLAU{For technical simplicity, we shall admit a continuum limit throughout, which in principle holds for distances 
$|x|=n\mathfrak{a}$ large enough with respect to the lattice constant $\mathfrak{a}$. 
On finite systems, this continuum assumption will work for lattice sizes $N$ large enough that the large-distance limit mentioned above can sensibly be taken.} 
As we shall show, the correlation function $C(t;x)$ 
on a semi-open lattice in the Glauber-Ising model corresponds to a the empty-interval probability $E(t,x)$ on a periodic ring in the
\BLAU{diffusion-coagulation process}, from which observables such as the concentration $c(t)$ can be found.

This paper is organised as follows. Section~2 recalls first the treatment of a finite, periodic chain in the Glauber-Ising model before the semi-open geometry is defined and then solved
through an extension of the spatial-symmetry method. We also discuss if/when a natural-looking short-cut towards the scaling function is applicable. 
In section~3  we shall show that the correlation function $C(t;x)$ 
of the semi-open Glauber-Ising model can be reinterpreted as the empty-interval probability $E(t,x)$ \BLAU{of diffusion-coagulation}
on a periodic ring. Conclusions are given in section~4. Three appendices contain technical details of the calculations. 

\section{Correlation functions on a finite chain} \label{sec:2}

We now describe the calculation of the spin correlation function which will be done throughout in the continuum limit such that $C_n(t) \mapsto C(t;x)$. 
We shall also restrict to $T=0$ since this is the only situation where dynamical scaling holds and slow dynamics occurs.

\subsection{\BLAU{Periodic chain}} \label{sec:2.1}
We begin by recalling the result in the $1D$ Glauber-Ising model on a periodic ring with $N$ sites \cite{Henk25b}. 
\BLAU{One purpose of this subsection is to briefly recall the technique we use here and we shall emphasise later the differences with respect to the semi-open chain.} 
In the continuum limit, one has from (\ref{1.4}) for the 
{correlation function} $C(t;x)$ the following equation of motion, together with the boundary conditions
\BEQ \label{2.1}
\partial_t C(t;x) = \partial_x^2 C(t;x) \;\; ; \;\; C(t;0) = C(t;N) = 1
\EEQ
and where $0\leq x\leq N$. 
The first of these constraints comes from the Ising constraint in (\ref{1.4}) and the second one comes from the periodicity. 
Rather than dealing with these directly, we recognise first that the physical correlation function $C(t;x)$ is even in $x$. This means that we can restrict to
positive value of $x$ only, such that $C(t;-x)$ with negative spatial arguments becomes available for purely mathematical purposes. 
This permits to treat the two constraints in (\ref{2.1}) by using an analytic continuation to negative values of $x$. From now on, $C(t;x)$ will denote that
analytically continued function and only at the very end, we revert to the physical correlation function by making the substitution $x \mapsto |x|$. 
The analytic continuation is expressed explicitly as follows \cite{Henk25b}
\BEQ \label{2.2}
C(t;-x) = 2 - C(t;x) \;\; , \;\; C(t;x) = C(t;N-x) ~~\Longrightarrow~~ C(t;x+2N)=C(t;x)
\EEQ
Clearly, the spatial symmetries (\ref{2.2}) reproduce the constraints in (\ref{2.1}). 
Together, these can be shown to imply that the analytically continued function $C(t;x)$ is periodic in the spatial coordinate $x$, with period $2N$, see (\ref{2.2}). 
In this way, the physically motivated constraints are embedded into spatial symmetry properties of the analytically continued function 
$C(t;x)$. Hence, for the analytically continued function, one has the Fourier series representation \cite{Hild06}
\BEQ \label{2.3}
C(t;x) = \sum_{k=-\infty}^{\infty} \wit{C}(t;k)\,e^{\II\pi k \frac{x}{N}} \;\; , \;\;
\wit{C}(t;k) = \frac{1}{2N} \int_{-N}^{N} \!\D x\: C(t;x) e^{-\II\pi k \frac{x}{N}}
\EEQ
and now the equation of motion (\ref{2.1}) can indeed be solved in Fourier space. A straightforward calculation leads for the physical correlation function to \cite{Henk25b}
\BEA
C(t;x) 
&=& \frac{1}{2N} \int_{0}^N \!\D x'\: \left\{ 2\vartheta_3\left(\frac{\pi}{2} \frac{|x|+x'}{N},e^{-\pi^2 t/N^2}\right) \right. \nonumber \\
& & \left. +C(0;x'\,) \left[ \vartheta_3\left(\frac{\pi}{2} \frac{|x|-x'}{N},e^{-\pi^2 t/N^2}\right) 
                           - \vartheta_3\left(\frac{\pi}{2} \frac{|x|+x'}{N},e^{-\pi^2 t/N^2}\right) \right] \right\}
\label{2.4}
\EEA
\BLAU{where $\vartheta_3$ is a Jacobi theta function \cite{Abra65}.} 
It follows that both the physical constraints as well as the required periodicity properties (\ref{2.2}) are indeed satisfied, if they only hold for the initial correlator $C(0;x)$. 
\BLAU{If the initial correlation function decays with $x$, for example $C(0;x)\sim |x|^{-\aleph}$ for large $|x|$ and with $\aleph>0$,} 
the term is irrelevant in the sense that it merely give\BLAU{s} rise to corrections to the leading finite-size scaling limit behaviour \BLAU{\cite{Henk04,Henk25b}. 
For example, for a fully disordered initial state with $C(0;x)\sim \delta(x)$, this correction term in the second line of (\ref{2.4}) vanishes.}.
\BLAU{The finite-size scaling behaviour is} obtained by taking simultaneously the limits $t\to\infty$, $|x|\to\infty$ and $N\to\infty$ but such that the (finite-size) scaling variables
\BEQ \label{2.5}
\mathfrak{u} = \frac{|x|}{t^{1/2}} \;\; , \;\; \mathfrak{v} = \frac{N}{t^{1/2}}
\EEQ
are kept finite. \BLAU{This scaling limit also} arises naturally for a fully disordered initial state\footnote{In general, an initial correlator
$C(0;x)$ in (\ref{2.4}) breaks dynamical scaling.} where $C(0;x)\sim \delta(x)$. \BLAU{Then one may} re-cast the physical single-time correlator (\ref{1.1}) as
\BEA 
C(t;x) &=& F_C^{\rm per}\left( \mathfrak{u}, \mathfrak{v} \right) = F_C^{\rm per}\left(\frac{|x|}{\sqrt{t\,}},\frac{N}{\sqrt{t\,}}\right) 
= \int_{0}^{1} \!\!\D u\: \vartheta_{3}\left(\frac{\pi}{2} u + \frac{\pi}{2}\frac{|x|}{N},e^{-\pi^2 t/N^2}\right) 
\nonumber \\
&=& 1 - \frac{2}{\pi}\int_{0}^{|x|/N} \!\!\!\D v\: \vartheta_2\left( \pi v,e^{-4\pi^2 t/N^2}\right) 
\label{2.6}
\EEA
\BLAU{where $\vartheta_{2}$ is another Jacobi theta function \cite{Abra65}, distinct from $\vartheta_3$.} 
This form is in generic agreement with the expectation of dynamical finite-size scaling \cite{Suzuki77}. 
The expressions (\ref{2.6}) give the explicit finite-size scaling functions for the single-time correlator, in terms
of the finite-size scaling variables $|x|/N$ and $t/N^2$, for a periodic chain\footnote{The generic form is quite analogous to existing analytical results in the
spherical model in $2<d<4$ dimensions, \BLAU{with periodic boundary conditions and} quenched to $T<T_c$ \cite{Henk23}, see also figure~\ref{fig2}(b) below.}  of length $N$. 

\begin{figure}[tb]
\begin{center}
\includegraphics[width=.45\hsize]{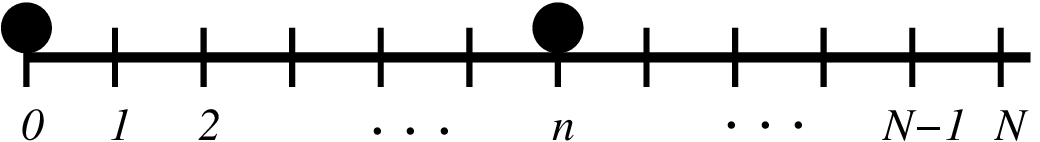}
\end{center}
\caption[fig1]{\small{Semi-open segment with $N$ sites. 
At the left, an Ising spin is kept fixed and the time-dependent correlator $C_n(t)$ with another spin at site $0<n<N$ is studied.
The constraints $C_0(t)=1$ and $C_N(t)=0$ are applied.}
\label{fig1} }
\end{figure}

\subsection{\BLAU{Semi-open chain}} \label{sec:2.2}
Our focus shall be on the semi-open chain, see fig.~\ref{fig1}. 
\BLAU{In contrast to the periodic ring considered in section~\ref{sec:2.1}, spatial translation-invariance no longer holds true.}
\BLAU{In one spatial dimensions, previous results with open boundary conditions include finite-size scaling studies on the critical relaxation times in the Glauber-Ising model \cite{Lusc96} 
or on the particle density in asymmetric exclusion models, via algebraic techniques, e.g. \cite{Derr92,Derr93,Schu98,Schu01}. 
To our knowledge, this is the first study of finite-size scaling, for free boundary conditions, on time-space correlation functions in the Glauber-Ising model.  
In figure~\ref{fig1},} we already used that the correlation function $C_n(t)=C_{-n}(t)$, \BLAU{should be considered as being of a central spin at fixed position $n=0$ and another
one $n$ sites away. This is symmetric} in $n$ and we can therefore consider $n\geq 0$ 
without restriction on the generality.\footnote{The only restriction we admit is that the site $n=0$ is fixed at the
centre of the interval $[-N,N]$.}  
\BLAU{Then we can also speak of the correlation function $C_n(t)=\langle \sigma_0(t) \sigma_n(t)\rangle$ between an Ising spin fixed at the leftmost edge of the interval $[0,N]$ 
and another spin at site $n$, to the right. We shall use this picture from now on.} 
The chain is open at the right end because of the requirement $C_N(t)=0$. In the continuum limit, we have  (with $0\leq x<N$) 
\BEQ \label{2.2.1}
\partial_t C(t;x) = \demi \partial_x^2 C(t;x) \;\; ; \;\; C(t;0) = 1 \;\; , \;\; C(t;N) = 0
\EEQ
\BLAU{Since the left spin is considered fixed, changes in $C(t;x)$ only arise from the motion of the right spin. 
This leads to a reduction of the diffusion constant by a factor 2 in comparison to the periodic case (\ref{2.1}) where both spins are mobile.\footnote{\BLAU{Consideration of two mobile
spins would require to use correlators with two spatial variables, in the spirit of \cite{Fort14}, but is beyond the scope of the present work.}}} 
Rather than dealing with these two constraints directly, we shall implement them via spatial symmetries in an analytically continued
function $C(t;x)$, in the same spirit as for the periodic case above. \BLAU{But in contrast to section~\ref{sec:2.1}, these} conditions are chosen to be
\BEQ \label{2.2.2} 
C(t;-x) = 2 - C(t;x) \;\; , \;\; C(t;x) = 1 - \frac{x}{N} + B(t;x) 
\EEQ
The first of these solves the first constraint. From the definition (\ref{2.2.2}) of the function $B(t;x)$,  the second constraint implies 
\BEQ \label{2.2.3}
B(t;\pm N) = B(t;0) = 0
\EEQ
In addition, combination with the first property (\ref{2.2.2}), 
proves that on the interval $[-N,N]$, the function $B$ is anti-symmetric (as shown in appendix~A)
\BEQ \label{2.2.4}
B(t;-x) = - B(t;x)
\EEQ
Since $B(t;\pm N)=0$ vanishes at the extremities of the interval $[-N,N]$ it can be considered to be of spatial period 
$2N$.\footnote{\BLAU{In spite of the absence of spatial translation-invariance, we have satisfied once more the conditions for a Fourier analysis to be applicable \cite[Kap. 4]{Hild06}.}}
Furthermore, it admits a Fourier representation 
\BEQ \label{2.2.5}
B(t;x) = \sum_{k=1}^{\infty} b_k(t) \sin\left(\frac{\pi}{N} k x\right) \;\; , \;\; 
b_k(t) = \frac{1}{N} \int_{-N}^{N} \!\D x\; B(t;x) \sin\left(\frac{\pi}{N} k x\right)
\EEQ
which is the analogue of (\ref{2.3}) above. Then the equation of motion (\ref{2.2.1}) can be solved in Fourier space and we find (see appendix~A for the details)
\BEA
C(t;x) &=& 1 - \int_0^{|x|/N} \!\D u\: \vartheta_3\left( \frac{\pi}{2} u, e^{-\frac{\pi^2}{2} \frac{t}{N^2}} \right) 
\label{2.2.6} \\
& & +\frac{1}{2N} \int_0^N \!\D x'\: C(0;x') \left[ 
\vartheta_3\left(\frac{\pi}{2}\frac{|x|-x'}{N}, e^{-\frac{\pi^2}{2} \frac{t}{N^2}} \right) 
- \vartheta_3\left(\frac{\pi}{2}\frac{|x|+x'}{N}, e^{-\frac{\pi^2}{2} \frac{t}{N^2}} \right) \right]
\nonumber
\EEA
where finally the analytically continued function is reduced to the physical correlation function $C(t;x)$ by making at the very end the substitution $x \mapsto |x|$. 
This gives the exact expression for the correlation function $C(t;x)$ between an Ising spin at the centre of the open segment  
$[-N,N]$ and another spin at the position $-N\leq x\leq N$,
subject to Glauber dynamics and such that the correlation function is forced to vanish $C(t;\pm N)=0$ at the end of the segment, see figure~\ref{fig1}.
Up to a trivial re-scaling in time (which comes from the rescaled equations of motion) 
the corrections to the leading scaling forms, for both the periodic and the semi-open lattice, are the same. \BLAU{It follows that the discussion of the irrelevance of spatially
decaying initial correlators can be taken over from the periodic case (\ref{2.4}) in section~\ref{sec:2.1} and hence also holds for the semi-open chain.}
The leading scaling contributions can be cast into their final forms
\begin{subequations} \label{2.2.7}
\begin{align}
C^{\rm semi}(t;x) &= 1 - \int_0^{|x|/N} \!\D v\: \vartheta_3\left( \frac{\pi}{2} v, e^{-\frac{\pi^2}{2} \frac{t}{N^2}} \right) \label{2.2.7a} \\
C^{\rm per}(t;x)  &= 1 - \frac{2}{\pi} \int_0^{|x|/N} \!\D v\: \vartheta_2\left( {\pi} v, e^{-{4\pi^2} \frac{t}{N^2}} \right)  \label{2.2.7b}
\end{align}
\end{subequations}
which depend on the finite-size scaling variables $|x|/N$ and $t/N^2$. 
They are both consistent with the generic expectations of dynamical finite-size scaling \cite{Suzuki77} 
in the sense that we can write the correlation functions (\ref{2.2.7}) as $C(t;x;N)=F_C\bigl(\frac{|x|}{\sqrt{t\,}}, \frac{N}{\sqrt{t\,}}\bigr)$ which generalises (\ref{1.1}) and where the
universal scaling functions $F_C$ are boundary-condition-dependent.  The existence of a second argument is the new feature of finite geometries. 

\begin{figure}[tb]
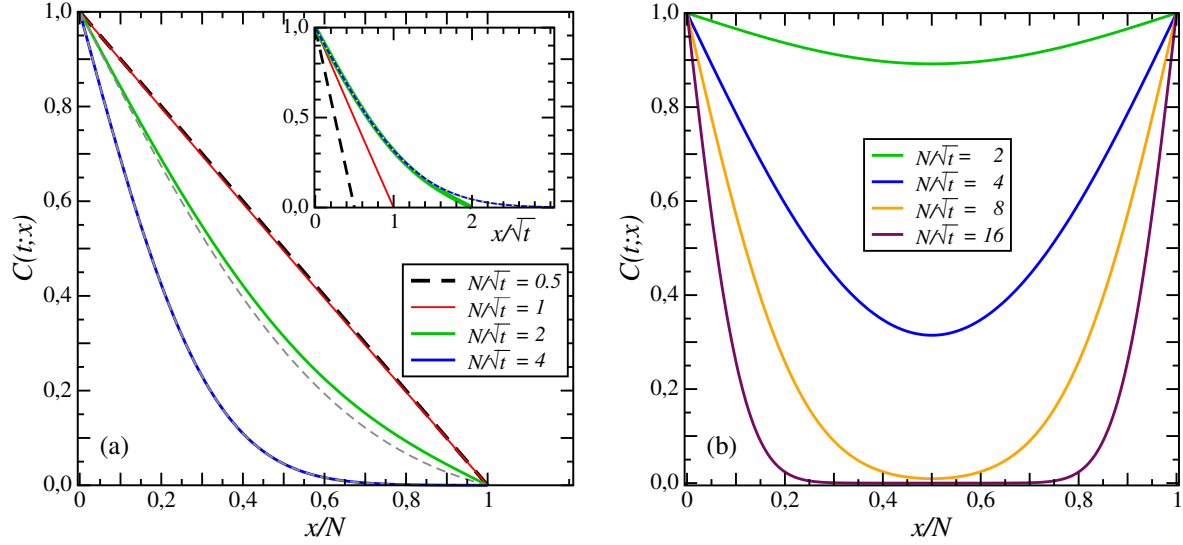

\begin{center}
\includegraphics[width=.44\hsize]{Glauber-semiouvert-correl-semiouvert-t10-skal.eps}~~~~\includegraphics[width=.44\hsize]{Glauber-semiouvert-correl-peri-t10-skal.eps}
\end{center}
\caption[fig2]{\small{Finite-size scaling in the $1D$ Glauber-Ising model at $T=0$ on a finite chain for (a) semi-open and (b) periodic boundary conditions. 
The main plots display the dependence of the correlation function $C(t;x)$ on the finite-size scaling variable $x/N$, for several fixed values of $N/\sqrt{t\,}$. The inset
in (a) shows the dependence of $C(t;x)$ on the bulk scaling variable $x/\sqrt{t\,}$, for the same values of $N/\sqrt{t\,}$. The dotted line in the inset is the infinite-size correlation
function $C_{\infty}(t;x)=\erfc(|x|/\sqrt{2t\,})$. The dashed gray lines in (a) give the approximate scaling function $C^{\rm scal}(t;x)$ according to (\ref{2.3.3}). 
}
\label{fig2} }
\end{figure}

Figure~\ref{fig2} illustrates these scaling functions: the semi-open correlator (\ref{2.2.7a}) in figure~\ref{fig2}(a) 
and the periodic correlator (\ref{2.2.7b}) in  figure~\ref{fig2}(b), over against the finite-size scaling variable $|x|/N$. 
A common aspect is that the functional forms of these correlators not only depend on the second finite-size scaling variable $N/\sqrt{t\,}$, 
but on the boundary conditions as well. At first sight, their behaviour is clearly quite distinct (notice that the qualitative shape of the curves in figure~\ref{fig2}(b) is similar to the
one found for the spherical model in $2<d<4$ dimensions, quenched \BLAU{to} temperature $T<T_c$ \cite{Henk23}). 
Upon closer inspection, it appears that the behaviour of the semi-open correlator (\ref{2.2.7a}) in the interval $0\leq \frac{|x|}{N}\leq 1$ is quite analogous, although not
identical, to the behaviour of the periodic correlator (\ref{2.2.7b}) in the interval $0\leq \frac{|x|}{N}\leq \demi$.\footnote{\BLAU{This also illustrates that the condition $C(t;N)=0$ does not
eliminate the possibilities of short-range order, analogously to the periodic case.}}  
Here the distinct boundary conditions $C^{\rm semi}(t;\pm N)=0$ and $C^{\rm per}(t;\pm N)=1$ are of essential influence. For example, at $x=N$, the semi-open correlator
vanishes exactly, whereas in the periodic case, at $x=\demi N$, the correlator tends to zero exponentially fast with increasing $N$. 
On  the other hand, if one considers, as in the inset of figure~\ref{fig2}a, the dependence of $C(t;x)$ on the bulk
scaling variable $x/\sqrt{t\,}$ the scaling functions are close the one of the spatially infinite system, viz. $C^{\rm semi}_{\infty}(t;x)=\erfc\bigl(\frac{|x|}{\sqrt{2t\,}}\bigr)$, 
if $N/\sqrt{t\,}$ is large enough.\footnote{Analogously, for periodic boundary conditions, in the limit $N/\sqrt{t\,}\to\infty$ the curves converge towards 
the correlator $C^{\rm per}_{\infty}(t;x)=\erfc\bigl(\frac{|x|}{2\sqrt{t\,}}\bigr)$ (not shown in figure~\ref{fig2}) \cite{Bray97}.} On the other side, if $\frac{|x|}{N}\ll 1$, one
finds a linear \BLAU{decay} whose slope \BLAU{approaches the one} of the infinite-size system, for $N/\sqrt{t\,}$ large enough. Deviations from the infinite-size curve become first
visible for large values of $\frac{|x|}{\sqrt{t\,}}$. 
\BLAU{Since the model is made from `hard' Ising spins, one finds a cusp at $x=0$, see figure~\ref{fig2}, usually referred to as
Porod's law \cite{Bray94a}. In contrast, such a cusp does not exist for `soft' spins, as they occur for example in the spherical model, 
where the correlator is rounded off at $x=0$ \cite[Fig. 1(a)]{Henk23}.}

\subsection{\BLAU{A short-cut towards dynamical scaling~?}} \label{sec:2.3}
Given that scaling approaches often allow for a rapid and simple derivation of (universal) scaling functions, it is tempting to try such a scaling
approach for the solution of the equation of motion (\ref{2.2.1}) and with its associated boundary conditions. 
One might try one's hand at a simple scaling {\em ansatz} of the form 
\BEQ \label{2.3.1}
C^{\rm scal}(t;x) = F\left( \frac{|x|}{\sqrt{t\,}}\right) \;\; ; \;\; \mathfrak{u} = \frac{|x|}{\sqrt{t\,}}
\EEQ
which should hold in the scaling limit where simultaneously $t\to\infty$, $x\to\infty$ but $\mathfrak{u}$ is kept finite \cite{Fort26}. Inserting the ansatz into
(\ref{2.2.1}) readily gives the differential equation
\BEQ
F''(\mathfrak{u}) + \mathfrak{u} F'(\mathfrak{u}) = 0 ~~\Longrightarrow~~ F(\mathfrak{u}) = F_0 + F_1 \int_0^{\mathfrak{u}} \!\D u'\; e^{-{u'}^2/2}
\EEQ
and where the two constants $F_{0,1}$ are to be found from the boundary conditions $F(0)=1$ and $F(N/\sqrt{t\,}\,)=0$, implied by (\ref{2.2.1}). This yields
\BEQ \label{2.3.3}
C^{\rm scal}(t;x) = 1 - \frac{\int_0^{x/\sqrt{t\,}} \!\D u'\: e^{-{u'}^2/2}}{\int_0^{N/\sqrt{t\,}} \!\D u'\: e^{-{u'}^2/2}} 
= 1 - \frac{\erf \frac{x}{\sqrt{2t\,}}}{\erf \frac{N}{\sqrt{2t\,}}}
= \frac{\erfc \frac{x}{\sqrt{2t\,}} - \erfc \frac{N}{\sqrt{2t\,}}}{1 - \erfc \frac{N}{\sqrt{2t\,}}}
\EEQ
where $\erf(x)$ is the error function \cite[(7.1.1)]{Abra65} and $\erfc(x)=1-\erf(x)$ is the complementary error function. 
Of course, such a scaling solution will be independent of any initial correlator $C(0;x)$. 
Since for large  arguments $\erfc(x)\stackrel{x\to\infty}{\to} 0$ exponentially fast \cite{Abra65}, 
the solution (\ref{2.3.3}) certainly has the attractive feature that for $N/\sqrt{t\,}\gg 1$ one recovers the
exactly known scaling function $C^{\rm scal}_{\infty}(t;x)\to \erfc\bigl( \frac{x}{\sqrt{2t\,}}\bigr)$ of the spatially infinite system \cite{Bray97}. 
\BLAU{We also observe that  the final result (\ref{2.3.3}) agrees once more with the generic expectations of dynamical finite-size scaling \cite{Suzuki77}, since it depends
on both variables $\frac{x}{\sqrt{t\,}}$ and $\frac{N}{\sqrt{t\,}}$, although from the original ansatz (\ref{2.3.1}) one might have expected a different result.}

\begin{figure}[tb]
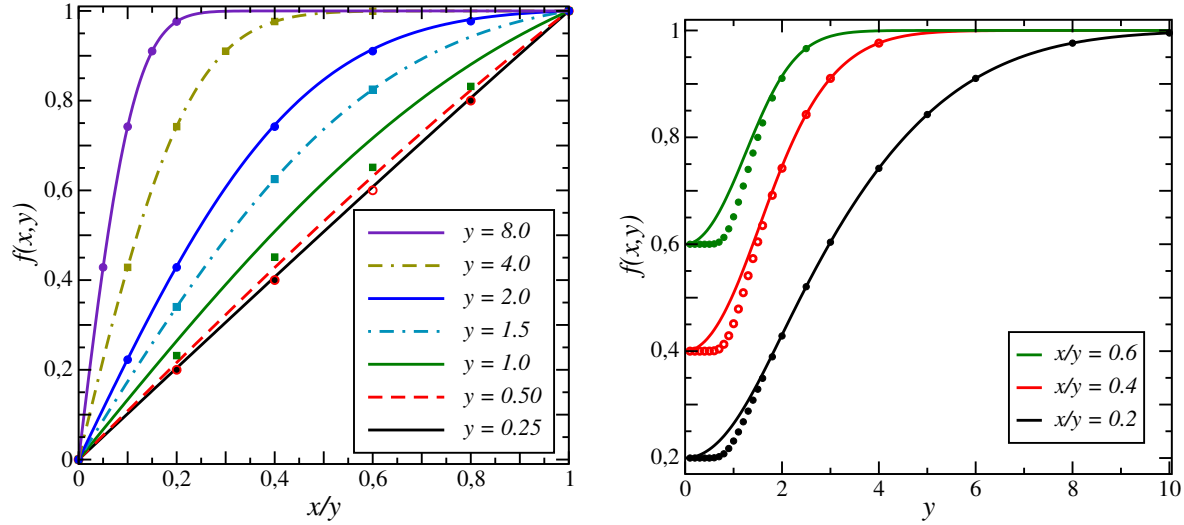

\begin{center}
\includegraphics[width=.44\hsize]{Glauber-semiouvert-ThetaInt-1.eps}~~~~\includegraphics[width=.44\hsize]{Glauber-semiouvert-ThetaInt-2.eps}
\end{center}
\caption[fig3]{\small{Test of the identity (\ref{2.3.4}), conjectured from the correlation scaling function of the finite-size semi-open $1D$ Glauber-Ising model. 
Full curves come from the simplified scaling expression derived from $C^{\rm scal}$. Points are derived from the exact solution $C^{\rm semi}$. 
The left panel shows the dependence on $x/y$ for several values of $y$. 
The right panel shows the dependence on $y$, for several vales of $x/y$.}
\label{fig3} }
\end{figure}

Does the approach leading to (\ref{2.3.3}) represent an useful short-cut in order to obtain the exact physical correlation function~? 
If that were so, comparison  with the exact solution (\ref{2.2.7a}) would imply the mathematical identity 
\BEQ \label{2.3.4} 
f(x,y) := \frac{\erf x}{\erf y} \stackrel{?}{=} \int_0^{x/y} \!\D u\: \vartheta_3\left( \frac{\pi}{2} u, e^{-\pi^2/4 y^2} \right) 
= \frac{2y}{\sqrt{\pi\,}} \int_0^{x/y} \!\D u\: e^{-y u^2/2}\, \vartheta_3\left( \II u y, e^{-4 y^2} \right)
\EEQ
where the last relation is a consequence of the modular transformation
\BEQ \label{2.3.5}
\vartheta_3\left( \pi u, e^{-\pi t}\right) 
= t^{-1/2} \exp\left(-\pi \frac{u^2}{t} \right) \vartheta_3\left( \II\pi \frac{u}{t},e^{-\pi/t}\right)
\EEQ
which follows from Poisson's re-summation formula \cite{Itzy89} of the Jacobi theta function 
$\vartheta_3\bigl(\frac{\pi}{2}u,q\bigr) = \vartheta_3\bigl(\frac{\pi}{2}(2-u),q\bigr)=\sum_{k\in\mathbb{Z}}q^{k^2}\cos(\pi u k)$ \cite[(16.27.3)]{Abra65}.  

In figure~\ref{fig3} numerical tests of the conjectured relation (\ref{2.3.4}) are shown. In the left panel, essentially the dependence on $x$ of the function $f(x,y)$ is displayed, for 
several values of $y$. The full curves show the left-hand side of (\ref{2.3.4}) as it follows from $C^{\rm scal}$, whereas the points show the right-hand side of (\ref{2.3.4}) as it
follows from $C^{\rm semi}$. Clearly, for $y$ large enough an excellent agreement is found and the points fall very nicely onto the full lines of the same colour. 
However, when $y\lesssim 2$, notable deviations appear, which are particularly notable around $y\approx 1$. The right panel illustrates this
further by showing $f(x,y)$ as a function of $y$, for several values of $x/y$. Again, for $y$ large enough, one observes a very good agreement 
(the points fall clearly onto the full lines of the same colour and deviations should be exponentially small)
but $C^{\rm scal}$ and $C^{\rm semi}$ lead to different results for small values of $y$, which appears to be most strong around $y\approx 1$. 
The same effect can also be seen in figure~\ref{fig2}(a), where the dashed gray lines give $C^{\rm scal}(t;x)$ for two values of $N/\sqrt{t\,}$. For the \BLAU{larger} one (blue curve), 
there is a very good agreement with the exact result (\ref{2.2.7a}). However, for the \BLAU{smaller} one (green curve), deviations are notable, although the curves are qualitatively similar. 

Hence the proposed identity (\ref{2.3.4}) only holds approximately, in the region $y\gtrsim 2$. 
The intriguing and simple short-cut towards a finite-size scaling function only produces an approximate
result, probably since the ansatz (\ref{2.3.1}) merely depends on the {\em single} scaling variable $\mathfrak{u}$. 
Comparison with the exact result (\ref{2.2.7a}) shows this to be an over-simplification \cite{Suzuki77}. 
\BLAU{Still, the quantitative agreement of the simple form (\ref{2.3.3}) with the exact result (\ref{2.2.7a}) is not so bad, see figure~\ref{fig2}a, 
and the simplicity of its derivation might become useful for a quick orientation in more complicated models.} 
Im\-pli\-cit\-ly, in the scaling approach described here, one has admitted that $N/\sqrt{t\,}\gg 1$ but figure~\ref{fig3} shows that the cross-over,
when the length scale $\ell(t)\sim \sqrt{t\,}$ becomes comparable to $N$, is not \BLAU{completely} captured. 

\section{Coagulation-diffusion process} \label{sec3}

The $1D$ Glauber-Ising model at temperature $T=0$ is dual \cite{Sigg77,Sant97} to the coagulation-diffusion process, of particles of a single species $A$ 
and provided diffusion $A+\emptyset \stackrel{D}{\longleftrightarrow} \emptyset+A$ and
coagulation $A+A \stackrel{D}{\longrightarrow} A+\emptyset, \emptyset+A$ occur with the same rate.\footnote{If the rates are different, the universal long-time exponent
$c(t)\sim t^{-1/2}$ is kept, but the associated amplitude will be modified, e.g. \cite{Odor01}. This is in agreement with experimental results, e.g. \cite{Shap18} and refs. therein.}  
In the exact solution, a central quantity is the empty-interval probability $E_n(t)$ \cite{benA90,Doer90,Krebs94a,Durang10,Durang11,Fort14}, 
which is the probability to find an interval of $n$ subsequent empty sites. Under the stated conditions, $E_n(t)$ obeys a closed set of equations of motion. 
In the continuum limit, one rather deals with a function $E(t,x)$ and which obeys the well-known equation of motion \cite{benA00,Krap10}
\BEQ \label{3.1} 
\partial_t E(t,x) = 2 D \partial_x^2 E(t,x) \;\; , \;\; E(t,0) = 1 \;\; , \;\; E(t,N) = 0
\EEQ
Herein, the last condition holds if the particles are moving on a ring of $N$ sites. 
If initially, there is at least one particle in the system, the last particle which has survived the
coagulation reactions cannot decay because of the lack of a reaction partner and on a ring of $N$ sites, the \BLAU{largest} empty interval can have $N-1$ sites. 
The other constraint follows since only nearest-neighbour particles can undergo a coagulation reaction. 
In what follows, we shall always scale to $D=1$. 

Clearly, the equations of motion (\ref{2.2.1}) and (\ref{3.1}), along with their constraints, are identical, 
up to a trivial re-scaling $t\mapsto 4t$ when going from the semi-open Glauber-Ising
model to the coagulation-diffusion process. On a periodic ring of size $N$, the empty-interval probability can be read off from the previous discussion
\BEA
E(t,x) &=& 1 - \int_0^{x/N} \!\D u\: \vartheta_3\left( \frac{\pi}{2} u, e^{-\frac{\pi^2}{2} \frac{4t}{N^2}} \right) 
\label{3.2} \\
& & +\frac{1}{2N} \int_0^N \!\D x'\: E(0,x') \left[ 
\vartheta_3\left(\frac{\pi}{2}\frac{x-x'}{N}, e^{-\frac{\pi^2}{2} \frac{4t}{N^2}} \right) 
- \vartheta_3\left(\frac{\pi}{2}\frac{x+x'}{N}, e^{-\frac{\pi^2}{2} \frac{4t}{N^2}} \right) 
\right]
\nonumber
\EEA
This is the precise statement of the duality with Glauber-Ising chain mentioned in section~\ref{sec1}, for the case of finite chains with $N$ sites. 
One of the quantities of interest is the time-dependent particle-concentration, which follows directly once $E(t,x)$ is known \cite{benA00,Krap10} and reads
\BEA
c(t) &=& - \left. \frac{\partial E(t,x)}{\partial x}\right|_{x=0} 
\nonumber \\
&=& \frac{1}{N} \vartheta_3\left( 0, e^{-2\pi^2 \frac{t}{N^2}} \right) 
\label{3.3} \\
& & +\left.\frac{1}{2N} \int_0^N \!\D x'\; E(0,x') \frac{\partial}{\partial x} \left( 
\vartheta_3\left( \frac{\pi}{2} \frac{x+x'}{N}, e^{-2\pi^2 t/N^2} \right) - \vartheta_3\left( \frac{\pi}{2} \frac{x-x'}{N}, e^{-2\pi^2 t/N^2} \right) \right)\right|_{x=0}
\nonumber 
\EEA
Herein, the first term does reproduce the well-known analytic result by Krebs {\it et al.} in their eq. (6.7) \cite{Krebs94a}. 
\BLAU{We point out that their result was derived on a discrete lattice (see their eq.~(6.6)) 
and a continuum limit was merely taken at the very end of their calculation (their solution also contains corrections
to scaling which come from the discreteness of the lattice)  \cite{Krebs94a}.} 
This \BLAU{observation} already serves as an useful cross-check of our calculational technique \BLAU{in section~\ref{sec:2.2}, 
including the use of the continuum limit right from the beginning.} 

The empty-interval probability can be expressed as
\BEQ
E(t,x) = \int_x^N \!\D x'\; P(t,x')
\EEQ
where $P(t,x)=\pr\bigl( \bullet\fbox{~$x$~}\,;t\bigr)$ is the probability to find an empty interval of size $x$ bounded on the left by a particle. Carrying out the partial integration
in (\ref{3.3}) leads to a more compact expression for the time-dependent density (as derived in appendix~B) 
\BEQ \label{3.5}
c(t) = \int_0^1 \!\D u\; P(0,N u)\, \vartheta_3\left(\frac{\pi}{2} u , e^{-2\pi^2 \frac{t}{N^2}} \right) 
\EEQ
If initially the particles are uncorrelated and have the infinite-volume concentration $c_{\rm eq}$, the initial probabilities on a finite ring may be chosen as 
\BEQ \label{3.6}
P(0,x) = \frac{c_{\rm eq}\, e^{-c_{\rm eq} x}}{1 - e^{-c_{\rm eq} N}} \;\; , \;\; E(0,x) = \frac{e^{-c_{\rm eq} x} - e^{-c_{\rm eq} N}}{1 - e^{-c_{\rm eq} N}}
\EEQ
The distribution $E(0,x)$ obeys the two constraints (\ref{3.1}), as it should. 
On a finite lattice of size $N$, $c_{\rm eq}=c_{\rm eq}(c_0,N)$ must be chosen such that the initial concentration 
takes indeed the desired value $c_0 = c_{\rm eq}/(1-e^{-c_{\rm eq}N})$, and consistent with $P(0,0)=c_0$. 
Physical arguments for this form for initially uncorrelated particles of concentration $c_0$ are recalled in appendix~C. 

Alternatively, one may also start from the initial distribution \cite{Fort26}
\BEQ \label{3.7}
P(0,x) = c_0 \left( 1 - \frac{x}{N}\right)^{c_0 N -1} \;\; , \;\; E(0,x) = e^{c_0 N \ln\bigl( 1 - x/N \bigr)}
\EEQ
which obeys the same boundary conditions as the choice (\ref{3.6}). Then the concentration can be found via
\BEQ
c(t) = c_0 \int_0^1 \!\D u\; u^{c_0 N -1}\, \vartheta_4\left(\frac{\pi}{2} u , e^{-2\pi^2 \frac{t}{N^2}} \right) 
\EEQ
with the theta function $\vartheta_4$ \cite{Abra65}. The scaling solution (the leading part in (\ref{3.3})) implicitly starts with an initial value $c_0=1$ 
and the modular transformation (\ref{2.3.5}) does produce the expected long-time decay $c(t) \simeq \bigl(2\pi\,t\bigr)^{-1/2}$. 
If an initial concentration $c_0<1$ is chosen, the concentration $c(t)$
will initially decay more slowly than the scaling solution and will cross over to the scaling decay once the more rapidly decaying scaling solution has become close to it.  

We have restricted attention here to the mere calculation of time-dependent concentrations $c(t)$. 
The study of many-point particle correlation functions should require the analysis of many-hole probabilities, e.g. following the lines of \cite{Durang10,Durang11,Fort14}. 

\section{Conclusions}

Even a century after its introduction \cite{Lenz20,Ising25}, 
and after a long history of having fruitfully stimulated many different insights into phase transitions at and far from equilibrium
(for historical reviews see \cite{Niss05,Berc09a,Berc09b,Giro20,Folk23,Folk26}),  
the {\it ``Ising model still thrives''} \cite{Fish81}. 
We have studied here some aspects of the celebrate Glauber-Ising dynamics \cite{Glau63} which in turn has become quite time-honoured itself. 
The Glauber-Ising dynamics in $1D$ is rightly famous, 
since it is one of the rare cases where the usually infinite hierarchy of coupled equations of motion \cite{Kreu86,Pott07} naturally decouples 
and thus becomes available to methods of analytical study. This feature has furnished explicit examples, in Ising model contexts, since a long time. 

In the continuum limit, time-space-dependent correlation functions $C(t;x)$ then obey simple diffusion equations but are still subject to boundary conditions
which prevent a totally straightforward solution, viz. in terms of Fourier analysis. Much of the numerous work in the past decades has been on how to treat these. 
Our main innovation in this work has been to show how to use spatial symmetries to recast the problem into one where Fourier series methods can indeed be used and then to show how this
applies in the case of non-periodic boundary conditions on finite lattices of size $N$. 
The results presented here can immediately be used as initial conditions 
for the calculation of two-time correlators and the exploration of novel finite-size effects therein \cite{Henk25c,Wark26}. 
We also hope that these techniques may become useful in different applications in the future, 
which may involve either more general interactions, \BLAU{surface magnetic fields} and/or more general boundary conditions. 
Similarly, in the related coagulation-diffusion process, it should be possible to consider particle currents at the boundaries \cite{Fort14} 
or to extend the techniques at hand towards the
analysis of correlation functions. Again, the boundary conditions which arise in the equations of motion for the empty-interval probabilities $E(t,x)$ were for a long time considered so 
difficult that attention was shifted to other observables which are more easy to analyse \cite{benA00,Krap10}. 

Explicit results were shown in figure~\ref{fig2} and satisfactorily enter into the generic and expected context of dynamical finite-size scaling \cite{Suzuki77}. 
This also provided the opportunity to test a proposal for a short-cut towards to the dynamical finite-size scaling functions, 
which although not exact \BLAU{still turns out to satisfy dynamical finite-size scaling and} might become of heuristic value in more complicated systems. 
Numerical work will now be needed to understand  further which aspects are specific to the $1D$ Glauber-Ising dynamics and which ones permit further generalisation. 
At the very least, our result should serve as a benchmark for future numerical studies. Long-standing relations with integrable quantum chains \cite{Alca94,Schu01} point towards possible
extensions towards quantum dynamics \cite{Malv21,Zahr26,Bouc26}. 

\newpage


\appsection{A}{Finite-size correlation function}
Details of the derivation of the correlation function (\ref{2.2.6}) are presented.  

We start from the equation of motion (\ref{2.2.1}) and its two accompagning constraints for the semi-open finite chain, see figure~\ref{fig1}. 
These are parametrised using (\ref{2.2.2}). Combining these leads to the condition
\BEQ
1 - \frac{-x}{N} + B(t;-x) = 2 - \left( 1 - \frac{x}{N} + B(t;x) \right) 
\EEQ
which upon simplification produces the anti-symmetry condition (\ref{2.2.4}). Hence the unknown function $B(t;x)$ is periodic on the 
interval $[-N,N]$ and can be cast into a Fourier series (\ref{2.2.4}) \cite{Hild06}. This implies for the Fourier coefficients $b_k(t)$
\BEQ
\partial_t {b}_k(t) = -\demi \left( \frac{\pi k}{N}\right)^2 b_k(t) ~~\Longrightarrow~~ b_k(t) = b_k(0) \exp\left[ -\demi \left( \frac{\pi k}{N}\right)^2 t\right]
\EEQ
and furthermore the integral representation
\BEQ
B(t;x) = \frac{1}{N} \int_{-N}^N \!\D x'\; B(0;x') 
         \sum_{k=1}^{\infty} \exp\left[ -\demi \left( \frac{\pi k}{N}\right)^2 t\right] \sin\left( \frac{\pi}{N} k x'\right) \sin\left( \frac{\pi}{N} k x\right) 
\EEQ
Since $C(t;x) = 1 - \frac{x}{N} + B(t;x)$, we find for the analytically continued correlator
\BEQ
C(t;x) = 1 - \frac{x}{N} + \frac{1}{N} \int_{-N}^N \!\D x'\; \left( C(0;x') - 1 +\frac{x'}{N} \right) 
\sum_{k=1}^{\infty} e^{ -\demi \left( \frac{\pi k}{N}\right)^2 t}\, \sin\left( \frac{\pi}{N} k x'\right) \sin\left( \frac{\pi}{N} k x\right) 
\EEQ
Since the last factor is odd in $x'$, the constant term in the parenthesis does not contribute to the integral. We then have the decomposition
\begin{subequations}
\begin{align}
C(t;x) &= 1 - \frac{x}{N} + \frac{1}{N} \int_{0}^N \!\D x'\; C(0;x')\, 
\sum_{k=1}^{\infty} e^{ -\demi \left( \frac{\pi k}{N}\right)^2 t}\, \sin\left( \frac{\pi}{N} k x'\right) \sin\left( \frac{\pi}{N} k x\right) 
\nonumber \\
&  ~~~+ \frac{1}{N} \int_0^{N} \!\D x'\; \underbrace{~C(0;-x')~}_{\BLAU{=} 2-C(0;x')}\, 
\sum_{k=1}^{\infty} e^{ -\demi \left( \frac{\pi k}{N}\right)^2 t}\, \sin\left( -\frac{\pi}{N} k x'\right) \sin\left( \frac{\pi}{N} k x\right) 
\nonumber \\
&  ~~~+ \frac{1}{N} \int_{-N}^N \!\D x'\; \frac{x'}{N} \sum_{k=1}^{\infty} e^{ -\demi \left( \frac{\pi k}{N}\right)^2 t}\, 
\sin\left( \frac{\pi}{N} k x'\right) \sin\left( \frac{\pi}{N} k x\right) 
\label{A.5a} \\
&= 1 - \frac{x}{N} - \frac{2}{N} \int_{0}^N \!\D x'\; 
\sum_{k=1}^{\infty} e^{ -\demi \left( \frac{\pi k}{N}\right)^2 t}\, \sin\left( \frac{\pi}{N} k x'\right) \sin\left( \frac{\pi}{N} k x\right) 
\nonumber \\
& ~~~-\frac{2}{N} \int_0^{N} \!\D x'\;C(0;x')\, \sum_{k=1}^{\infty} e^{ -\demi \left( \frac{\pi k}{N}\right)^2 t}\: 
\demi \left[ \cos\left(\frac{\pi}{N}(x'+x)k\right) - \cos\left(\frac{\pi}{N}(x'-x)k\right) \right] 
\nonumber \\
& ~~~+ \frac{1}{N} \int_{-N}^N \!\D x'\; \frac{x'}{N} \sum_{k=1}^{\infty} e^{ -\demi \left( \frac{\pi k}{N}\right)^2 t}\, 
\sin\left( \frac{\pi}{N} k x'\right) \sin\left( \frac{\pi}{N} k x\right) 
\label{A.5b} \\
&=: 1 - \frac{x}{N} + T_1 + T_2 + T_3
\label{A.5c}
\end{align}
\end{subequations} 
where the first continuation (\ref{2.2.2}) is applied to the second line of (\ref{A.5a}) and \cite[(4.3.31)]{Abra65} was used in the second line of (\ref{A.5b}). 

With the help of the identity 
\BD
\int_0^1 \!\D u\: \sin\bigl( \pi k u\bigr) = \left\{ 
\begin{array}{ll} 0 & \mbox{\rm\small ~;~~ $k$ even} \\ \frac{2}{\pi k} & \mbox{\rm\small ~;~~ $k$ odd} \end{array} \right.
\ED
the first term in (\ref{A.5c})  becomes
\BEA
T_1 &=& - \frac{2}{N} \int_{0}^N \!\D x'\; 
\sum_{k=1}^{\infty} e^{ -\demi \left( \frac{\pi k}{N}\right)^2 t}\, \sin\left( \frac{\pi}{N} k x'\right) \sin\left( \frac{\pi}{N} k x\right) 
\nonumber \\
&=& -2  \sum_{k=1}^{\infty} e^{ -\demi \left( \frac{\pi k}{N}\right)^2 t}\, \sin\left( \frac{\pi}{N} k x\right) \int_0^1 \!\D u\: \sin\bigl( \pi k u\bigr) 
\nonumber \\
&=& - \frac{4}{\pi} \sum_{k=0}^{\infty} e^{ -\frac{\pi^2 t}{2 N^2} \bigl(2k+1\bigr)^2}\, \frac{1}{2k+1}\sin\left( \frac{\pi x}{N} \bigl(2k+1\bigr)\right) 
\EEA
Next, the identity 
\BD
\int_{-1}^{1} \!\D u\; u \sin\bigl( \pi k u\bigr) = - \frac{2}{\pi k} (-1)^k
\ED
gives for the third term
\BEQ
T_3 =  \sum_{k=1}^{\infty} e^{ -\demi \left( \frac{\pi k}{N}\right)^2 t}\, \sin\left( \frac{\pi}{N} k x\right) 
\frac{1}{N} \int_{-N}^N \!\D x'\; \frac{x'}{N} \sin\left( \frac{\pi}{N} k x'\right)
=- \sum_{k=1}^{\infty} e^{ -\frac{\pi^2 t}{2N^2} k^2}\, \frac{2}{\pi k} (-1)^k \sin\left( \frac{\pi x}{N} k\right) 
\EEQ
such that both terms together turn into 
\BEA
\lefteqn{ T_1 + T_3 = -\frac{4}{\pi} \sum_{k=0}^{\infty} e^{ -\frac{\pi^2 t}{2 N^2} \bigl(2k+1\bigr)^2}\, \frac{1}{2k+1}\sin\left( \frac{\pi x}{N} \bigl(2k+1\bigr)\right) 
              - \frac{2}{\pi} \sum_{k=1}^{\infty} e^{ -\frac{\pi^2 t}{2N^2} k^2}\, \frac{(-1)^k}{k} \sin\left( \frac{\pi x}{N} k\right)} 
\nonumber \\
&=& \left(- \frac{4}{\pi} +\frac{2}{\pi}\right) \sum_{k=0}^{\infty} \frac{e^{ -\frac{\pi^2 t}{2 N^2}\bigl(2k+1\bigr)^2}}{2k+1}\sin\left( \frac{\pi x}{N} \bigl(2k+1\bigr)\right)
- \frac{2}{\pi} \sum_{k=1}^{\infty} \frac{e^{ -\frac{\pi^2 t}{2N^2} \bigl(2k\bigr)^2}}{2k} \sin\left( \frac{\pi x}{N} 2k\right)
\nonumber \\
&=& - \frac{2}{\pi} \sum_{k=1}^{\infty} e^{ -\frac{\pi^2 t}{2N^2} k^2}\, \frac{1}{k} \sin\left( \frac{\pi x}{N} k\right)
\EEA
where in the first line, the second sum is decomposed into odd and even integers and in the second line a partial cancellation arises, before the two sums can be gathered together again. 
With the identity $\int_0^x \!\D x'\; \cos\bigl( \frac{\pi x'}{N} k\bigr) = \frac{N}{\pi k} \sin\bigl( \frac{\pi k}{N} x\bigr)$, this becomes
\BEQ \label{A.9}
T_1 + T_3 = -\frac{2}{\pi} \frac{\pi}{N} \int_0^x \!\D x'\; \sum_{k=1}^{\infty} e^{ -\frac{\pi^2 t}{2N^2} k^2}\, \cos\left( \frac{\pi x'}{N} k\right) 
= \frac{x}{N} - \frac{1}{N} \int_0^x \!\D x'\;  \vartheta_3\left( \frac{\pi}{2} \frac{x'}{N}, e^{-\frac{\pi^2}{2 N^2} t} \right) 
\EEQ
and with the Jacobi theta function $\vartheta_3$ \cite[(16.27.3)]{Abra65}. 
This contribution is independent of the initial correlator. Finally, the second term
\BEA
\lefteqn{T_2 = -\frac{1}{N} \int_0^{N} \!\D x'\;C(0;x')\, \sum_{k=1}^{\infty} e^{ -\demi \left( \frac{\pi k}{N}\right)^2 t}
         \left[ \cos\left(\frac{\pi}{N}(x'+x)k\right) - \cos\left(\frac{\pi}{N}(x'-x)k\right) \right]} 
\nonumber \\
&=& -\frac{1}{2N} \int_0^N \!\D x'\; C(0;x') \left[ 
\vartheta_3\left( \frac{\pi}{2} \frac{x+x'}{N}, e^{-\pi^2 t/2 N^2} \right) 
- \vartheta_3\left( \frac{\pi}{2} \frac{x-x'}{N}, e^{-\pi^2 t/2 N^2} \right) \right]~~
\label{A.10}
\EEA
can be expressed in terms of the Jacobi theta function $\vartheta_3$ as well. 
In particular, $T_2$ vanishes for an uncorrelated initial condition $C(0;x)\sim \delta(x)$. 

Insertion of the results (\ref{A.9},\ref{A.10}) into (\ref{A.5c}) does produce (\ref{2.2.6}) in the text, where
we also substituted $x\mapsto |x|$ in order to retrieve the physical correlation function. 

\appsection{B}{Particle-density in the coagulation-diffusion process}

Eq.~(\ref{3.5}) is derived. This is based on the known empty-interval probability (\ref{3.2}) and the fact that the average
time-dependent concentration can be found as the derivative $c(t)=\left.-\partial_x E(t,x)\right|_{x=0}$ \cite{benA00}. 

We begin by considering the contribution related to the initial probability in (\ref{3.2}). Standard trigonometric identities give 
\BEA
\lefteqn{ \left. \frac{\partial}{\partial x} \left( 
\vartheta_3\left( \frac{\pi}{2} \frac{x-x'}{N}, e^{-2\pi^2 t/N^2}\right) 
- \vartheta_3\left( \frac{\pi}{2} \frac{x+x'}{N}, e^{-2\pi^2 t/N^2}\right) 
\right) \right|_{x=0} }
\nonumber \\
&=& 2 \left. \sum_{k=1}^{\infty} e^{-\bigl(2\pi^2 t/N^2\bigr) k^2} \frac{\partial}{\partial x} \left( 
\cos\left( \frac{\pi}{N}(x-x')k\right) - \cos\left( \frac{\pi}{N}(x-x')k\right) \right) \right|_{x=0} 
\nonumber \\
&=& 4 \left. \sum_{k=1}^{\infty} e^{-\bigl(2\pi^2 t/N^2\bigr) k^2} \frac{\partial}{\partial x} \left(
\sin\left( \frac{\pi}{N} x k\right) \sin\left( \frac{\pi}{N} x' k\right) \right) \right|_{x=0}
\nonumber \\
&=& 4 \frac{\pi}{N} \left. \sum_{k=1}^{\infty} e^{-\bigl(2\pi^2 t/N^2\bigr) k^2}
    k \underbrace{~\cos\left( \frac{\pi}{N} x k\right)~}_{=1} \sin\left( \frac{\pi}{N} x' k\right) \right|_{x=0}
\EEA
This is then inserted into the calculation of the concentration
\BEA
\lefteqn{ c(t) = \left. - \frac{\partial E(t,x)}{\partial x}\right|_{x=0} }
\nonumber \\
&=& \frac{1}{N} \vartheta_3\left(0, e^{-2\pi^2 t/N^2}\right) 
  - \frac{2\pi}{N^2} \int_0^N \!\D x'\; E(0,x')  \sum_{k=1}^{\infty} e^{-\bigl(2\pi^2 t/N^2\bigr) k^2}\, k \sin\left( \frac{\pi}{N} x' k\right)
\nonumber \\
&=& \frac{1}{N} \vartheta_3\left(0, e^{-2\pi^2 t/N^2}\right) 
+ \frac{2}{N} \sum_{k=1}^{\infty} e^{-\bigl(2\pi^2 t/N^2\bigr) k^2} 
\int_0^N \!\D x'\; E(0,x') \left( \frac{\partial}{\partial x'} \cos\left( \frac{\pi}{N} x' k\right) \right)
\nonumber \\
&=& \frac{1}{N} \vartheta_3\left(0, e^{-2\pi^2 t/N^2}\right) \nonumber \\
& & + \frac{2}{N} \sum_{k=1}^{\infty} e^{-\bigl(2\pi^2 t/N^2\bigr) k^2} \left\{ \left.E(0,x') \cos\left( \frac{\pi}{N} x' k\right)\right|_0^N
-\int_0^N \!\D x'\; \frac{\partial E(0,x')}{\partial x'} \cos\left( \frac{\pi}{N} x' k\right) \right\} ~~~~~
\EEA
after a partial integration. With the boundary conditions $E(0,0)=1$, $E(0,N)=0$ of the empty-interval probability, the concentration becomes
\BEA
c(t) &=& \frac{1}{N} \vartheta_3\left(0, e^{-2\pi^2 t/N^2}\right) 
       + \frac{2}{N} \sum_{k=1}^{\infty} e^{-\bigl(2\pi^2 t/N^2\bigr) k^2} 
       \left\{ -1 -\int_0^N \!\D x'\; \frac{\partial E(0,x')}{\partial x'} \cos\left( \frac{\pi}{N} x' k\right) \right\}
\nonumber \\
&=& \frac{1}{N} \left( 1 + 2 \sum_{k=1}^{\infty} e^{-\bigl(2\pi^2 t/N^2\bigr) k^2} 
                         - 2 \sum_{k=1}^{\infty} e^{-\bigl(2\pi^2 t/N^2\bigr) k^2} \right) 
\nonumber \\
& & -\frac{1}{N} \int_0^N \!\D x'\; \frac{\partial E(0,x')}{\partial x'}  
\left( -1 + 1 + 2\sum_{k=1}^{\infty} e^{-\bigl(2\pi^2 t/N^2\bigr) k^2} \cos\left( \frac{\pi}{N} x' k\right) \right) 
\nonumber \\
&=& \frac{1}{N} + \frac{1}{N} \int_0^N \!\D x'\; \frac{\partial E(0,x')}{\partial x'}  
-\frac{1}{N} \int_0^N \!\D x'\; \frac{\partial E(0,x')}{\partial x'} \vartheta_3\left(\frac{\pi}{2}\frac{x'}{N}, e^{-2\pi^2 t/N^2}\right) 
\label{B.3} 
\EEA
To evaluate the second integral further, we recall that
\BEQ
E(t,x) = \int_x^N \!\D x'\; P(t,x') \;\; ; \;\; 
P(t,x) = \pr\bigl( \bullet\fbox{~$x$~}\,;t) = \begin{array}{l} \mbox{\rm\small probability of empty section of size $x$,}\\
\mbox{\rm\small bounded on the left by a particle}\end{array}
\EEQ
hence $P(t,x) = -\partial_x E(t,x)$. The usefulness of this quantity was pointed out by Durang {\it et al.} \cite{Durang10}. 
The first integral in (\ref{B.3}) is treated using the same boundary conditions as above. This finally implies
\BEA
c(t) &=& \frac{1}{N} +\frac{1}{N} \bigl( E(0,N) - E(0,0) \bigr) 
+ \frac{1}{N} \int_0^N \!\D x'\; P(0,x')\, \vartheta_3\left(\frac{\pi}{2}\frac{x'}{N}, e^{-2\pi^2 t/N^2}\right) 
\nonumber \\
&=& \frac{1}{N} \int_0^N \!\D x'\; P(0,x')\, \vartheta_3\left(\frac{\pi}{2}\frac{x'}{N}, e^{-2\pi^2 t/N^2}\right) 
\EEA
which is the assertion stated in the text. 

\appsection{C}{Initial states in the coagulation-diffusion process}

Physical arguments to justify the use of eq.~(\ref{3.6}) are recalled. 

This is done in the context of particles hopping freely on a lattice \cite{benA00}. 
At equilibrium, one has a state of maximal entropy and particle-distribution is random (Poissonian). 
It may be described in terms of the so-called {\em interparticle distribution function} ({\sc ipdf}) $p_{\rm eq}(x)$. 
A Poissonian distribution is exponential $p_{\rm eq}(x)=c e^{-cx}$ \cite{benA00} with the particle concentration $c$. 
On the other hand, for finite times $t$, the {\sc ipdf} $p(t,x)$ is related to the empty-interval probability $E(t,x)$, namely \cite{benA00,Krap10}
\BEQ
c(t) p(t,x) = \frac{\partial^2 E(t,x)}{\partial x^2} \;\;  , \;\; P(t,x) = -\frac{\partial E(t,x)}{\partial x} = \pr\bigl( \bullet\fbox{~$x$~}\,;t)
\EEQ
where $c(t)$ is the time-dependent concentration and $P(t,x)$ the probability to find an empty interval of size $x$, bounded on the left by a particle. 
Clearly, $c(t)p(t,x) = -\partial_x P(t,x)$. 

If one has at equilibrium a Poissonian distribution, encoded via $p_{\rm eq}(x)=c_{\rm eq}\, e^{-c_{\rm eq}x}$, 
it is clear that $P_{\rm eq}(x)=c_{\rm eq}\, e^{-c_{\rm eq}x}$ is Poissonian as well. 
For a spatially infinite system, which for large times relaxes towards equilibrium with $c(t)\to c_{\rm eq}$, one has
\BEQ
E(\infty,x) = E_{\rm eq}(x) = \int_x^{\infty} \!\D x'\; P_{\rm eq}(x') = \int_{x}^{\infty} \!\D x'\: c_{\rm eq} e^{-c_{\rm eq} x} = e^{-c_{\rm eq} x}
\EEQ
reproducing the well-known form for independent particles of concentration $0<c_{\rm eq}\leq 1$ \cite{benA90,Krap10}. 

\begin{figure}[tb]
\begin{center}
\includegraphics[width=.4\hsize]{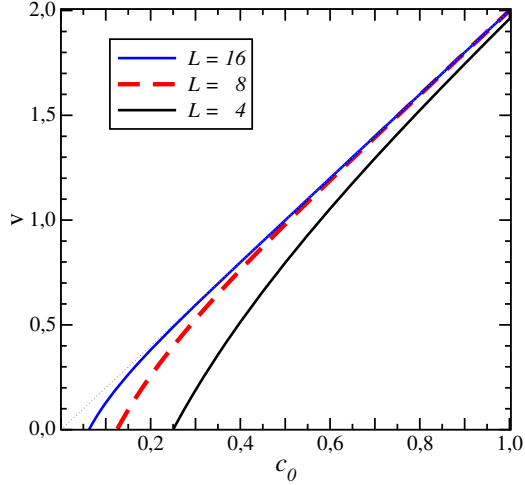}
\end{center}
\caption[figC1]{\small{Choice of the parameter $v=v(c_0,L)$ 
required for achieving the desired initial concentration $c_0$, for $L=[4,8,16]$ from bottom to top. 
}
\label{figC1} }
\end{figure}

On a finite chain of length $L$, at equilibrium the {\sc ipdf} is expected to be of the form 
\BEQ
p_{\rm eq}(x) = p_0\, c_{\rm eq}\, e^{-c_{\rm eq} x}
\EEQ
and should be normalised according to $\int_0^L \!\D x\; p_{\rm eq}(x)\stackrel{!}{=} 1$. This fixes $p_0$ such that
\BEQ 
p_{\rm eq}(x) = \frac{c_{\rm eq}\, e^{-c_{\rm eq} x}}{1 - e^{c_{\rm eq} L}} = P_{\rm eq}(x)
\EEQ
and hence
\BEQ
E_{\rm eq}(x) = \int_x^{L} \!\D x'\; P_{\rm eq}(x') = \frac{c_{\rm eq}}{1-e^{-c_{\rm eq} L}} \int_x^L \!\D x'\; e^{-c_{\rm eq} x} 
= \frac{e^{-c_{\rm eq} x} - e^{-c_{\rm eq} L}}{1 - e^{-c_{\rm eq} L}}
\EEQ
which is (\ref{3.6}) in the text. 

One may also obtain this form via reversible coagulation-decoagulation which besides single-particle diffusion $A+\emptyset \longleftrightarrow \emptyset +A$ also contains
the reversible reactions $A+A \longleftrightarrow A$ \cite{Doer90}. Therein, the empty-interval probability $E(t,x)$ obeys the equation
\BEQ
\partial_t E(t,x) = 2 \partial_x^2 E(t,x) + v \partial_x E(t,x)
\EEQ
with the decoagulation rate $v$. 
At equilibrium, one has $E(t,x)\stackrel{t\to\infty}{\longrightarrow} E_{\rm eq}(x)$ and $\partial_t E_{\rm eq}(x)=0$. 
With the boundary conditions $E_{\rm eq}(0)=1$ and $E_{\rm eq}(L)=0$, one readily obtains \cite{Doer90,benA90}
\BEQ \label{C.7}
E_{\rm eq}(x) = \frac{e^{-\frac{v}{2} x} - e^{-\frac{v}{2} L}}{1 - e^{-\frac{v}{2} L}}
\EEQ
The state (\ref{3.6}) may be achieved as follows: begin with the reversible coagulation-decoagulation process and relax it towards equilibrium, 
choosing $v=v(c_0,L)$ by solving $c_0 = \frac{v}{2}/\bigl(1 - e^{-v L/2}\bigr)$ for $v$, such as to obtain the desired concentration $c_0$, see figure~\ref{figC1}. 
Having fixed this configuration, the rate $v$ for decoagulation $A\longrightarrow 2A$ is set to zero and the state (\ref{C.7}) so prepared is taken as the initial state, 
at time $t=0$, for the subsequent irreversible coagulation-diffusion process. 

Alternatively, one may also consider the statistics of a hole of $n$ sites on a lattice of size $L$. This leads to the initial configuration \cite{Fort26}
\BEQ
E(0,x) = E(0,x) = e^{c_0 N \ln\bigl( 1 - x/N \bigr)} \;\; , \;\; P(0,x) = -\frac{\partial E(0,x)}{\partial x} = c_0 \left( 1 - \frac{x}{N}\right)^{c_0 N -1}
\EEQ
where $c_0=P(0,0)$ is the initial concentration. 
This satisfies the same boundary conditions as the initial configuration prepared above and is stated in (\ref{3.7}) in the text. 


\noindent
{\bf Acknowledgements:} It is a pleasure to thank J.-Y. Fortin for useful discussions. 
This work was supported by the french ANR-PRME UNIOPEN (ANR-22-CE30-0004-01).  


\end{document}